\documentclass[preprint,showpacs,preprintnumbers,amsmath,amssymb]{revtex4}

\usepackage{graphicx}
\usepackage{dcolumn}
\usepackage{bm}

\def\bit{\begin{itemize}}
\def\eit{\end{itemize}}

\def\beq{\begin{equation}}
\def\eeq{\end{equation}}
\def\bea{\begin{eqnarray}}
\def\eea{\end{eqnarray}}
\begin{document}
\preprint{FNAL-PUB-02/152-E;physics/0207083}
\title{A Measure of the Goodness of Fit in Unbinned Likelihood Fits}
\author{Rajendran Raja}
 \email{raja@fnal.gov}
\affiliation{%
Fermi National Accelerator laboratory\\
Batavia, IL 60510}

\date{\today}

\begin{abstract}

Maximum likelihood fits to data can be done using binned data
(histograms) and unbinned data. With binned data, one gets not only
the fitted parameters but also a measure of the goodness of fit. With
unbinned data, currently,  the fitted parameters are obtained but
no measure of  goodness of fit is available. This remains, to date, an
unsolved problem in statistics. Using Bayes theorem and likelihood
ratios, we provide a method by which both the fitted quantities and a
measure of the goodness of fit are obtained for unbinned likelihood
fits, as well as errors in the fitted quantities. We provide an ansatz for 
determining Bayesian {\it a priori} probabilities.
\end{abstract}

\maketitle

\section{Introduction} 
We outline a method by which goodness of fit
measures can be calculated in an unbinned likelihood analysis. We are
able to also calculate the probability density function of the fitted
variables and hence their errors in a rigorous manner.  We briefly
describe the currently used method of ``maximum likelihood'',
originally due to R.A.Fisher~\cite{fisher}. 
Let $s$ denote a set parameters defining
our theoretical model used to describe data.  Example of $s$ are the
mass of the top quark or  the lifetime of a particle. The symbol  
$s$ (for signal) can in general denote a discrete or continuous set of
variables. Let $c$ denote a set of observations describing a high 
energy physics event
and there are $n$ events in our dataset. In general, for each event,
$c$ can be a vector of dimension $d$.  Let $P(c|s)dc$ describe the
probability of observing the configuration $c$ in the $d$-dimensional
phase space volume $dc$ given the theoretical parameter set $s$. Thus
$P(c|s)$ is a probability density function ($pdf$) in the variable $c$
and obeys
\begin{equation}
\int P(c|s) dc = 1
\end{equation}
Then one can define a likelihood $\cal L$ of observing the dataset as
\begin{equation}
{\cal L} = \prod_{i=1}^{i=n} P(c_i|s)
\end{equation}
 The maximum likelihood point can be found of observing by minimizing 
the negative log-likelihood $-log_e \cal
 L$ defined as
\begin{equation}
-log_e{\cal L} = -\sum_{i=1}^{i=n} log_e(P(c_i|s))
\label{mx1}
\end{equation}
while varying the parameters $s$ either analytically or numerically
 to obtain the best values $s^*$ of $s$ that fit the data.

At the maximum likelihood point, $s^*$, the best fit values of $s$, 
are obtained. There is however no measure of the goodness of fit, since the
likelihood at the optimal value is not normalized to anything. There
is strictly no measure for the error on the fitted parameters, since
$\cal L$ is not a probability density function of $s$, though people
have calculated errors by treating the $-log_e \cal L$ at the minimum
as though it were equivalent to $\frac{1}{2}\chi^2$. Such error
calculations are not hitherto considered rigorously justifiable.

Unbinned likelihood fits, despite these disadvantages, are extremely
useful in finding $s^*$ since one does not have to treat bins with
small populations in a special manner as would be the case for binned fits.

In this paper we use Bayes theorem to rectify the above
disadvantages. In the process, we obtain a measure for the goodness of
fit and also  $P(s|c)$, the posterior
$pdf$ of $s$, enabling us to calculate the errors of the fitted
values in a rigorous way.

\section {Bayes Theorem}
We derive Bayes theorem here for the sake of completeness and to
illustrate the main ideas.  In the Bayesian approach \cite{bayes}, the
theoretical parameters $s$ can have a probability distribution both
{\it a priori} and {\it a posteriori}. The {\it a priori} distribution
refers to the knowledge of $s$ before the given set of observations
are made. The {\it a posteriori}  probability
distribution refers to the distribution of $s$, given the set of
observations $c$.
\subsection{Joint and Conditional Probabilities}
 We define a joint probability density for the theory parameters $s$
 and the observables $c$ as
\begin{equation}
        dP_{joint} = P_{joint}(s,c) ds\,dc
\end{equation}
 which is the probability that $s$ occurs in interval $s$ and $s+ds$
 and $c$ occurs in a volume element $dc$ centered around $c$.

We define the conditional probability density
\begin{equation}
        dP_{conditional} = P(s|c) ds
\end{equation}
as the probability density of observing $s$ in the interval $s$ and
$s+ds$ {\it given} that $c$ occurs in a volume element $dc$ centered
around $c$ .

Similarly, the conditional probability density
\begin{equation}
        dP_{conditional} = P(c|s) dc
\end{equation}
is defined as the probability density of observing $c$ in a volume
element $dc$ centered around $c$, {\it given} that $s$ occurs between
$s$ and $s+ds$ .
Then, by the laws of probability , we can write the joint probability
\begin{equation}
        dP_{joint} = P_{joint}(s,c) ds \, dc = P(c|s)dc \times P(s)ds
\label{eq1}
\end{equation}
Where $P(s)$ is the {\it a priori} probability of observing $s$ in
interval $s$ and $s+ds$ , and $P(c|s)dc$ is the probability of
observing $c$ given $s$ .  One can also obtain the same joint
probability, by first observing $c$ with {\it a priori} probability
$P(c)$ and then using the conditional probability $P(s|c)$ , i.e.
\begin{equation}
        dP_{joint} = P_{joint}(s,c) ds dc = P(s|c)ds \times P(c)dc
\label{eq2}
\end{equation}
By equating \ref{eq1} and \ref{eq2}, one gets the fundamental relation
leading up to Bayes Theorem.
\begin{equation}
    dP_{joint} = P(c|s)dc \times P(s)ds = P(s|c)ds \times P(c)dc
\end{equation}
Expressed in terms of densities, dropping $ds$ and $dc$ terms, this
yields
\begin{equation}
 P(c|s) \times P(s) = P(s|c) \times P(c)
\label{eq33}
\end{equation}
 One is interested in evaluating $P(s|c)$, the probability of the
 theory parameters, given a set of observations $c$.
This becomes,
\begin{equation}
P(s|c) = \frac{ P(c|s) \times P(s)}{P(c)}
\label{eq3}
\end{equation}
The {\it a priori} probability $P(c)$ is not an independent quantity,
given the {\it a priori} probability $P(s)$ which represents the
knowledge of $s$ before the set of observations $c$ . The reason for
this is that $P(s|c)$ integrated over $s$ must add up to unity.

\subsection{Some Normalization Formulae}
Integrating over one of the variables in the joint probability yields,
using equation \ref{eq1}, the following relations.
\begin{equation}
P(c) \equiv \int P_{joint}(s,c) ds = \int P(c|s) \times P(s)ds
\end{equation}
where the $\equiv$ sign is the definition of the {\it a priori}
probability $P(c)$, since one integrates the joint probability
$P_{joint}(s,c)$ over all values of $s$ . This then yields
\begin{equation}
P(c) = \int P(c|s) \times P(s)ds
\label{normpc}
\end{equation}
also, integrating the joint probability over $c$, one gets
\begin{equation}
P(s) \equiv \int P_{joint}(s,c) dc = \int P(c|s) \times P(s)dc
\end{equation}
i.e.
\begin{eqnarray}
P(s) = \int P(c|s) \times P(s)dc \\
\:or \int P(c|s) dc = 1
\end{eqnarray}
Similarly, using equation \ref{eq2}, one gets relations similar to the
above with $c$ and $s$ interchanged. Summarizing, one gets the
following normalization relations.
\begin{eqnarray}
P(c) = \int P(c|s) \times P(s)ds \label{eqn1}\\ P(s) = \int P(s|c)
\times P(c)dc \label{eqn2}\\
\int P(s) ds = 1  \\
\int P(c) dc = 1\label{eqn2.1}  \\
\int P(c|s) dc = 1 \label{eqn3}\\
\int P(s|c) ds = 1 \label{eqn4}
\end{eqnarray}
Substituting \ref{eqn1} in \ref{eq3}, one gets the derivation of Bayes
Theorem.
\begin{equation}
P(s|c) = \frac{P(c|s)\times P(s)}{ \int P(c|s) \times P(s)ds}
\label{eqn5}
\end{equation}
The above equation normalizes to unity as per equation \ref{eqn4}.  This is the
central expression of Bayes' theorem.

\subsection{Observation of Many Configurations}
Now we come to one of the more beautiful properties of formula
\ref{eqn5}, namely it is recursive. Let us observe two separate
configurations say, $c_1$ and $c_2$ . Then equation \ref{eqn5} yields
for $c_1$ ,
\begin{equation}
P(s|c_1) = \frac{P(c_1|s)\times P(s)}{ \int P(c_1|s) \times P(s)ds}
\label{eqn7}
\end{equation}
Now we observe $c_2$. We wish to compute $P(s|c_1,c_2)$ , the
probability of $s$ given $c_1$ and $c_2$ .  We can then replace the
{\it a priori} probability for $s$, $P(s)$ in equation \ref{eqn5} by
the probability of $s$ after observing $c_1$ (i.e. $P(s|c_1)$) to
calculate the probability of $s$ given $c_1$ and $c_2$ . This yields,
\begin{equation}
P(s|c_1,c_2) = \frac{P(c_2|s)\times P(s|c_1)}{ \int P(c_2|s) \times
P(s|c_1)ds}
\label{eqn10}
\end{equation}
Substituting for $P(s|c_1)$ from equation \ref{eqn7}, we get
\begin{eqnarray}
P(s|c_1,c_2) = \frac{P(c_2|s)P(c_1|s)P(s)/ \int P(c_1|s)P(s)ds} {\int
P(c_2|s)P(c_1|s)P(s)ds / \int P(c_1|s)P(s)ds} \\ \mbox{yielding}
~P(s|c_1,c_2) = \frac{P(c_2|s)P(c_1|s)P(s)} {\int
P(c_2|s)P(c_1|s)P(s)ds} \\ \mbox{generalizing,}~P(s|c_1,c_2...c_n) =
\frac{P(c_n|s)...P(c_2|s)P(c_1|s)P(s)} {\int
P(c_n|s)...P(c_2|s)P(c_1|s)P(s)ds}
\label{eqn8}
\end{eqnarray}
Another way to think about equation \ref{eqn8} is to think of the $n$
configurations as one massive super configuration $\bf c_n$ , which
also obeys the Bayes theorem equation \ref{eqn5}
\begin{eqnarray}
P(s|{\bf c_n}) = \frac{P({\bf c_n}|s)\times P(s)} { \int P({\bf
c_n}|s) \times P(s)ds}\label{eqnbay} \\ \mbox{where} ~ P({\bf c_n}|s) =
P(c_n|s)...P(c_2|s)P(c_1|s)\label{eqnind}
\end{eqnarray}
It should be noted that the probability $P({\bf c_n}|s)$ obeys the
normalization condition \ref{eqn3}. Equation~\ref{eqnind} is just the
law of multiplication of independent probabilities. This implies that
it is possible to chain probabilities in Bayes theorem as in
equation~\ref{eqn10} if and only if the configurations are
statistically independent. This is certainly true in the case of high
energy physics events.

The expression for {\it a posteriori} probability $P(s|{\bf c_n})$ in
equation~\ref{eqnbay} cannot be used as is unless one knows P(s), the
{\it a priori} probability of $s$. In the ``Bayesian approach'',
people use various guesses for
P(s) and a lot of care and energy are expended in arriving at
``reasonable'' functions for P(s).

\subsection{Likelihood Ratios}

We now recast the Bayes theorem equation~\ref{eqnbay} as a set of
likelihood ratios ${\cal L_R}$.

\begin{equation}
{\cal L_R} = \frac{P(s|{\bf c_n})}{P(s)} = \frac{P({\bf
c_n}|s)}{P({\bf c_n})}
\label{eqnlr}
\end{equation}
where we have substituted the function $P(\bf{c_n})$ for the
normalizing integral in the denominator using equation~\ref{normpc}.
The likelihood ratio ${\cal L_R}$ has a very important invariant
property.  It is invariant under the transformations of variable sets
$c\rightarrow c'$ and $s\rightarrow s'$ where $c'$ and $s'$ are
functions of the variable sets $c$ and $s$.
It is possible to ask what exact variables one uses to form the
vector $c$. For instance,when a jet is measured experimentally, 
does one use the energy, pseudo-rapidity and
azimuth of  the jet or the three components of the energy three vector as components of $c$?  
Clearly, the
probability density function $P(c|s)$ will depend on the choice of the
variable set $c$ since,
\begin{equation}
 P(c'|s) = |\frac{dc}{dc'}| P(c|s)
\end{equation}
where, $|\frac{dc}{dc'}|$ denotes a Jacobian of transformation to go
from the set of variables $c$ to $c'$.  However, the same Jacobian
ocurs in the denominator of the ${\cal L_R}$, hence the likelihood
ratio is unaffected by the transformation. The same argument can be
made with respect to transformations of the variable set $s\rightarrow
s'$. These are extremely important properties, so we henceforth work
with the likelihood ratio ${\cal L_R}$ and not the likelihoods ${\cal
L}$ which do not possess these properties.
\section{The Principle of Maximum Likelihood Ratios}
The equation~\ref{eqnlr} for ${\cal L_R}$ can be expanded as follows.
\begin{equation}
{\cal L_R} = \frac{P({\bf c_n}|s)}{P(\bf{c_n})} =
\frac{P(c_1|s)}{P(c_1)}\times\frac{P(c_2|s)}{P(c_2)}\dots
\times\frac{P(c_n|s)}{P(c_n)}
\label{eqnbay1}
\end{equation}
where we have used the independence of {\it a priori} probabilities
for $P(c_i),i=1,n$.  Similarly, one gets expressions,
\begin{equation}
{\cal L_R} = \frac{P(s|{\bf c_n})}{P(s)} =
\frac{P(s|c_1)}{P(s)}\times\frac{P(s|c_2)}{P(s)}\dots
\times\frac{P(s|c_n)}{P(s)}
\label{eqnbay2}
\end{equation}
where we have derived equation~\ref{eqnbay2} from
equation~\ref{eqnbay1} by applying equation~\ref{eqnlr} to the
likelihood ratios of the individual events in the product.
In order to find the optimal set of parameters $s$, we  maximize
the likelihood ratio ${\cal L_R}$ in equation~\ref{eqnbay1} with
respect to $s$.  This is equivalent to minimizing the negative log
likelihood ratio $log_e{\cal L_R}$.
\begin{equation}
-\frac{\partial log_e {\cal L_R}}{\partial s} = -\sum_{i=1}^{i=n}
\frac{\partial log_e P(c_i|s)}{\partial s}=0
\label{maxlr}
\end{equation}
Notice that this is the same set of equations that one gets when
maximizing the likelihood as in equation~\ref{mx1}, since the {\it a
priori} probabilities $P(c_i)$ are constant with respect to variations
in $s$. So one gets the same set of optimal variables $s^*$ whether
one maximizes the likelihood ${\cal L}$ or the likelihood ratio $\cal
L_R$.  However, at the optimum, the likelihood ratio can be used to
obtain a goodness of fit parameter as we show below, whereas the
likelihood method would be unable to provide this information.
One can now ask what the minimum value of ${\cal L_R}$ is with respect
to variations in the event configuration, for a fixed value of theory;
i.e. what event configurations produce the minimum value of the
negative log likelihood?  Differentiating equation~\ref{eqnbay1} with
respect to $c_i$, one gets,
\begin{equation}
-\frac{\partial log_e {\cal L_R}}{\partial c_i} = -
\frac{\partial log_e P(c_i|s)}{\partial c_i} +
\frac{\partial log_e P(c_i)}{\partial c_i} =0
\label{maxlr1}
\end{equation}
i.e
\begin{eqnarray}
\frac{\partial log_e P(c_i|s)}{\partial c_i} =
\frac{\partial log_e P(c_i)}{\partial c_i}\\
P(c_i|s) = P(c_i)
\label{maxlr2}
\end{eqnarray}
The equation~\ref{maxlr2} implies that the lowest value of the likelihood ratio
occurs when the experimental probability density $P(c)$ and the theory
probability density $P(c|s)$ are the same at the observed events. The negative
log likelihood is zero at this point, yielding the best possible fit.

\section{Evaluating the Function $P(\bf c)$ and the Goodness of Fit}
 The key point to note is that just as $P(s)$ is the {\it a priori}
probability of the theoretical parameter $s$, $P(c)$ is the {\it a priori}
probability of the data.
In order to evaluate the likelihood ratio ${\cal L_R}$ at the maximum
likelihood point, one needs to evaluate the function $P(c)$ at the
observed event configurations $c_1,c_2\dots c_n$.  So the problem to
solve is this: given the event configurations $c_1,c_2\dots c_n$, what is
their probability density? Well known methods exist to estimate 
the $pdf's$ given discrete event distributions. These are collectively titled
probability density estimators~($PDE$), which have recently 
found application in high energy physics analyses~\cite{pde}. 

As noted above, the probability density function $P(c)$ is the {\it a
priori} $pdf$ of the data. In previous applications, to the author's
best knowledge, the function $P(c)$ was subsumed into the
equation~\ref{normpc} and expressed in terms of an unknown
$P(s)$. This resulted in the theory $pdf$ $P(c|s)$ being evaluated at
the data points $c_1,c_2\dots c_n$, but not the data $pdf$!  It is
precisely this failure to evaluate $P(c)$ given $c$ that has led to
the absence of goodness of fit criteria in unbinned likelihood fits.

In binned likelihood fits, one fits a theoretical curve to a binned
set of data points. Two distributions, those of theory and data,
are involved in providing a goodness 
of fit measure such as $\chi^2$ in the binned approach.
In the unbinned method, however,
one finds the maximum likelihood point by evaluating the theoretical
function $P(c|s)$ at the data points $c_i, i=1,n$. There is only one
distribution involved, namely theory! One has hitherto ignored $P(c)$,
by subsuming it into a normalization constant. We rectify this lapse
here and restore $P(c)$ to its proper role, namely, the $pdf$ of the data.

\subsection{Probability Density Estimators}
Let $c^\alpha _i,\:\alpha=1,d$ denote the components of the 
d-dimensional vector $c$ for the i$^{th}$ event. 
Then we can define the mean vector $<c^\alpha>$ as
\begin{equation}
 <c^\alpha> = \frac{1}{n} \sum_{i=1}^{i=n} c^\alpha_i
\end{equation}
The covariance (or error) matrix $E$ of $c$ is defined as
\begin{equation}
 E^{\alpha,\beta} = <c^\alpha c^\beta > -<c^\alpha><c^\beta>
\end{equation}
where the $<>$ implies average over the $n$ events. The Hessian matrix $H$ 
is defined as the inverse of $E$.
One can define a  multivariate Gaussian Kernel ${\cal G} (c)$
as
\begin{equation}
 {\cal G}(c) = \frac{1}{(\sqrt{2\pi}h)^d\sqrt(det(H))}
exp(\frac{-H^{\alpha\beta}c^\alpha c^\beta}{2h^2})
\end{equation}
where the repeated indices imply summing over and the parameter $h$ is a 
``smoothing parameter'', 
which has\cite{hoptim} a suggested optimal value $h\approx n^{-1/(d+4)}$.
The $pdf$ of $c$ is then given by
\begin{equation}
  P(c)\approx PDE(c) = \frac{1}{n}\sum_{i=1}^{i=n} {\cal G}(c-c_i)
\label{pde}
\end{equation}
Simply put, one takes an arbitray point $c$ in configuration space,
calculates the separation from this point to all the measured points
and sums up the values (at $c$) of the Gaussians that are centered at
the measured points.  This sum is divided by the number of Gaussians,
which equals $n$.  Since the Gaussians are all normalized to unity,
$P(c)$ obeys equation~\ref{eqn2.1}. One can feed in any value of $c$
and the $PDE$ will provide a probability density at that value of $c$.
It is clear that the $PDE$ method is generalizable to arbitrary
dimensions and will provide us with $P(c)$. One should note that the
Gaussian Kernel function depends on $n$, the number of events in the
sample. This dependence is through the smoothing parameter, which goes
to zero as $n\rightarrow\infty$. In this limit, equation~\ref{pde}
becomes
\begin{equation}
 P(c) = \int P(c) G_\infty(c-c_i)dc_i
\end{equation}
This implies that 
\begin{equation}
  G_\infty(c-c_i)\equiv\lim_{n \rightarrow \infty} G(c-c_i) = \delta (c-c_i)
\end{equation}
There exist generalizations~\cite{towers} of the above scheme where
the covariance matrix is made locally variable that can estimate
$pdf's$ with greater complexity albeit at a cost in computing
speed. In what follows, we introduce a method by which the smoothing
factor can be made a function of the configuration variables $c$ in an
iterative fashion, which may be equivalent to varying the covariance
matrix locally.

\section{An Illustrative Example}
\begin{figure}[tbh!]
\centerline{\includegraphics[width=4.0in]{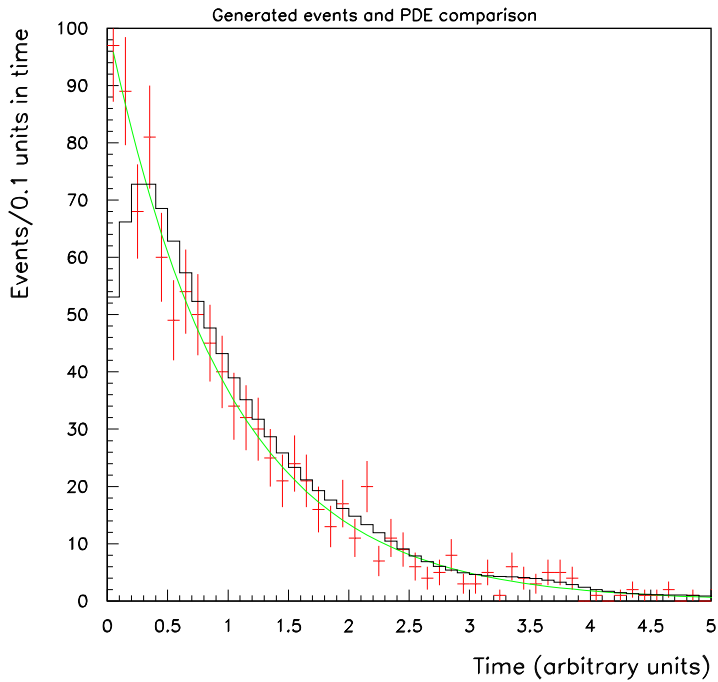}}
\caption[generated events]
{Figure shows the histogram (with errors) of generated events. 
Superimposed is the theoretical
curve $P(c|s)$  and the $PDE$ estimator (solid) histogram with no errors.
\label{genev}}
\end{figure}

We illustrate the ideas discussed above with a simple one-dimensional
example of events in which the observable $c$ consists of decay times
distributed exponentially with a decay constant $s$=1.0 in arbitrary
units.  The conditional probability $P(c|s)$ defines our theoretical
model and is given by
\begin{equation}
P(c|s) = \frac{1}{s}\exp(-\frac{c}{s})
\end{equation}
The $PDE$ one dimensional Gaussian kernel for this simple case would be
given by
\begin{equation}
{\cal G}(c) = \frac{1}{(\sqrt{2\pi} s h)}
\exp (-\frac{c^2}{2s^2h^2})
\end{equation}
We generate a thousand events for which the smoothing parameter $h$ is
calculated to be 0.125 as per equation\cite{fudge} $h=0.5 n^{-1/(d+4)}$.
Figure~\ref{genev} shows the generated events, the theoretical curve
$P(c|s)$ and the $PDE$ curve $P(c)$ normalized to the number of
events.
The $PDE$ curve is a poor estimator of the data near the cutoff at
$c$=0. This is because the Gaussians centered at values of negative
$c$ would have contributed to the curve near $c$=0. Also, for large
values of $c$, data are sparse and the Gaussian approximation with
constant smoothing factor $h$ finds it difficult to approximate the
data. We choose to restrict our fitting to a fiducial interval in time
$t_1<~c~< t_2$ =
$1<~c~<5$. Both the theoretical model $P(c|s)$ and the $PDE$ likelihood
curves are renormalized so that they integrate to unity in the
fiducial interval.

\subsection{Iterative Determination of the Smoothing Factor}

The expression $h\approx n^{-1/(d+4)}$ clearly is meant to give a
smoothing factor that decreases slowly with increased statistics
$n$. It is expected to be true on average over the whole 
distribution. However, the exponential distribution under consideration
has event densities that vary by orders of magnitude as a function of
the time variable $c$. In order to obtain a function $h(c)$ that takes
into account this variation, we first work out a $PDE$ with
constant $h$ and then use the number densities obtained thus~\cite{power} 
to obtain $h(c)$ as per the equation
\begin{equation}
  h(c) =   \left(\frac{n\:PDE(c)}{(t_2-t_1)}\right)^{-0.6}
\end{equation}

The equation is motivated by the consideration that a uniform
distribution of events between $t_1$ and $t_2$ has a $pdf= 1/(t_2-t_1)$
whereas the real $pdf$ is approximated by $PDE$. The function
$h(c)$ thus obtained is used to work out a better $PDE(c)$. This process 
is iterated three times to give the best smoothing function.

\begin{figure}[tbh!]
\centerline{\includegraphics[width=4.0in]{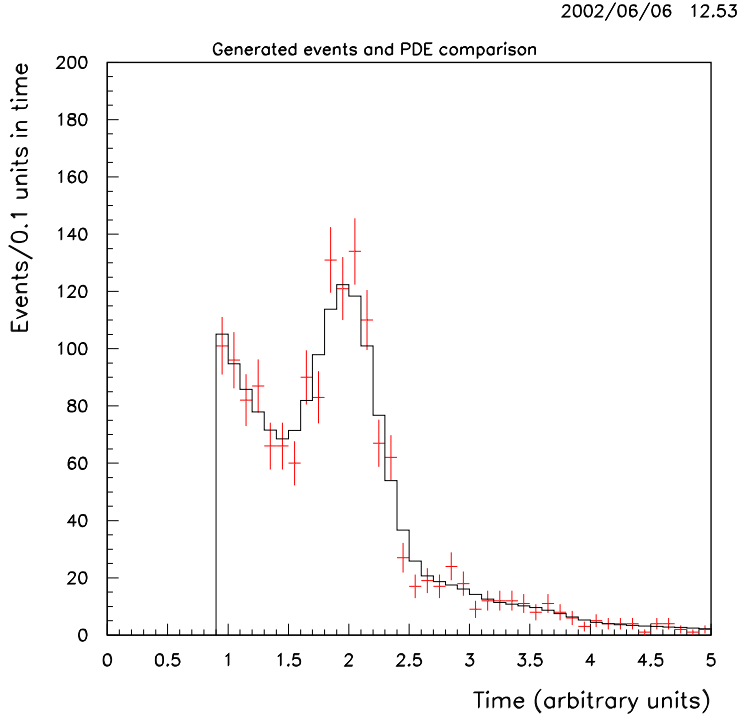}}
\caption[generated events]
{Figure shows the histogram (with errors) of 1000 events in the 
fiducial interval $1.0<c<5.0$ generated as
an exponential with decay constant $s$=1.0. with a superimposed
Gaussian of 500 events centered at $c$=2.0 and width=0.2.  
The $PDE$ estimator is
the (solid) histogram with no errors. 
\label{genev1}}
\end{figure}

We generate $n$=1000 events in the fiducial interval. If now we
were to superimpose a Gaussian with 500 events centered at $c$=2.0 and
width=0.2 on the data, the $PDE$ estimator will follow the data as
shown in Figure~\ref{genev1}. This shows that the $PDE$ estimator we have 
is adequate to reproduce the data, once the smoothing parameter is 
made to vary with the number density appropriately.

\begin{figure}[tbh!]
\centerline{\includegraphics[width=4.0in]{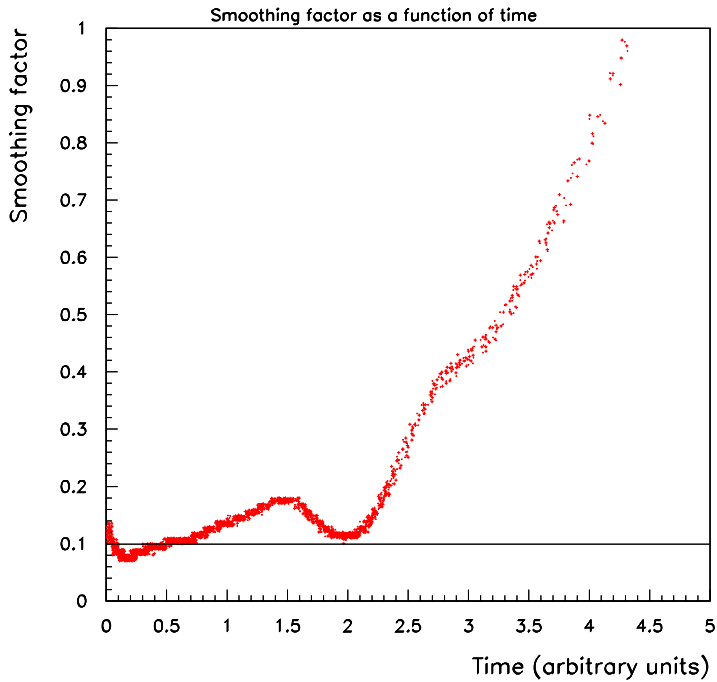}}
\caption[smoothing factor as a function of $c$]
{The variation of $h$ as a function of $c$ for the example shown in 
Figure~\ref{genev1}. The variation of the smoothing parameter is obtained 
iteratively as explained in the text. The flat curve is a smoothing 
factor resulting from the formula $h\approx 0.5n^{-1/(d+4)}$.
\label{hc}}
\end{figure}

The smoothing function $h(c)$ for the events in Figure~\ref{genev1}
is shown in Figure~\ref{hc}.  It can be seen that the value of $h$
increases for regions of low statistics and decreases for regions of
high statistics. Superimposed is the constant smoothing factor
obtained by the equation $h \approx 0.5 n^{-1/(d+4)}= 0.5n^{-0.2}$,
with $n$ being the total number of events generated, including those
outside the fiducial volume.

\subsection{An Empirical Measure of the Goodness of Fit}

\begin{figure}[tbh!]
\centerline{\includegraphics[width=4.0in]{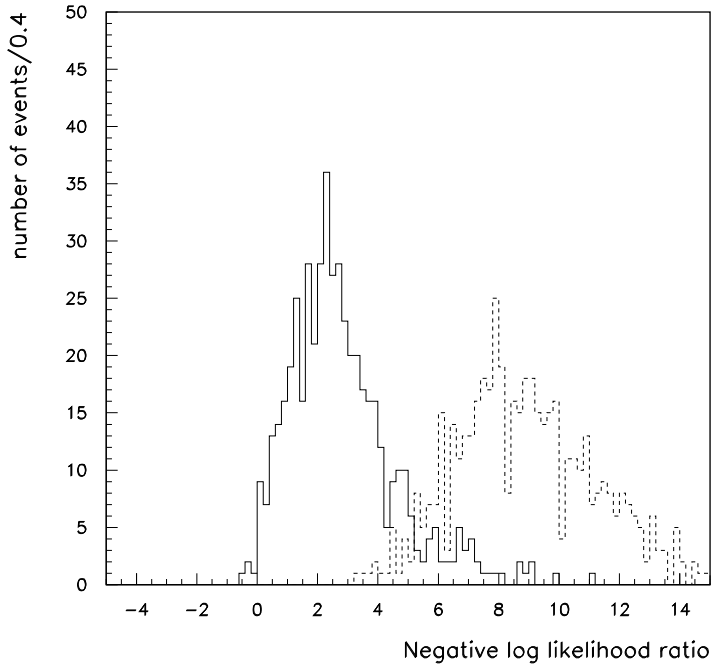}}
\caption[Distribution of fitted negative log-likelihood ratios]
{The solid curve shows the distribution of the negative log likelihood
ratio ${\cal NLLR}$ at the maximum likelihood point for 500
distributions, using the iterative smoothing function mechanism. The
dashed curve shows the corresponding distribution in the case of a
constant smoothing function.
\label{fitlike}}
\end{figure}

\begin{figure}[tbh!]
\centerline{\includegraphics[width=4.0in]{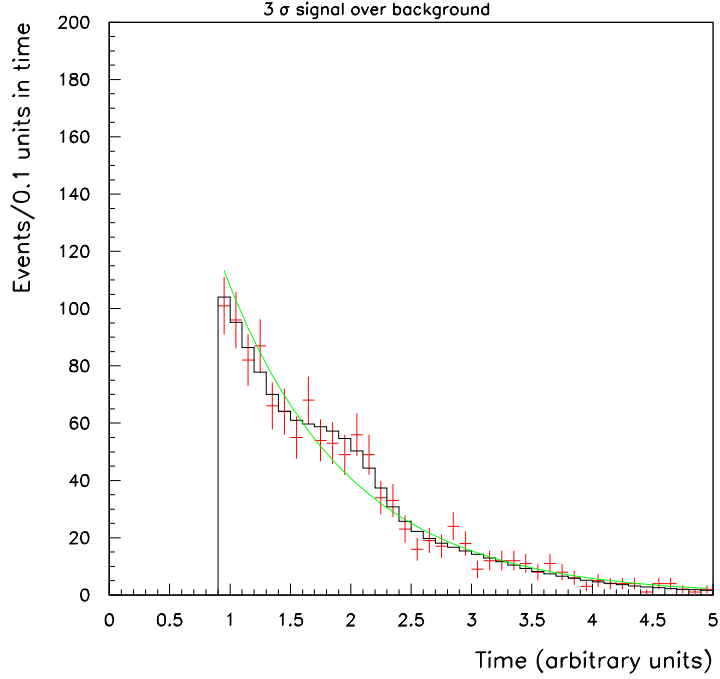}}
\caption[fitted events 3$\sigma$ case]
{Figure shows the histogram (with errors) of 1000 events in the 
fiducial interval $1.0<c<5.0$ generated as
an exponential with decay constant $s$=1.0. with a superimposed
Gaussian of 85 events centered at $c$=2.0 and width=0.2.  
The $PDE$ estimator is
the (solid) histogram with no errors. The data are fitted with a goodness 
of fit that is 3$\sigma$ away from the average value of ${\cal NLLR}$.
The continuous curve shows the fit to an exponential.
\label{genev2}}
\end{figure}

The negative log-likelihood ratio ${\cal NLLR} \equiv -{log_e {\cal
L_R}}$ at the maximum likelihood point now provides a measure of the
goodness of fit. In order to use this effectively, one needs an analytic theory
of the sampling distribution of this ratio. This is difficult to
arrive at, since this distribution is sensitive to the smoothing
function used. If adequate smoothing is absent in the tail of the
exponential, larger and broader sampling distributions of ${\cal
NLLR}$ will result. 
One can however determine the distribution of ${\cal NLLR}$
empirically, by generating the events distributed according to the
theoretical model many times and determining ${\cal NLLR}$ at the
maximum likelihood point for each such distribution. The solid
histogram in figure~\ref{fitlike} shows the distribution of ${\cal
NLLR}$ for 500 such fits.  This has a mean of 2.8 and an $rms$ of 1.8. The
dotted histogram shows the corresponding value of ${\cal NLLR}$ for
the constant value of smoothing factor shown in figure~\ref{hc}. This
distribution is clearly broader ($rms$=2.63) with a higher mean(=9.1) 
and thus has less discrimination power in judging the goodness of fit 
than the solid curve.
We now fit the same exponential background distribution with different
numbers of Gaussian events superimposed on an exponential background.
Table~\ref{tab1} shows the results of the fit. When a Gaussian of 500
events with width 0.2 and mean 2.0 is superimposed on the exponential
distribution of 1000 events, a value of ${\cal NLLR}$=189 is obtained
while trying to fit for the exponential using the unbinned maximum
likelihood method. This is 103$\sigma$ away from the mean of the
${\cal NLLR}$ distribution shown in figure~\ref{fitlike} with the
iterated smoothing function. A 3$\sigma$ effect is observed when the
number of events in the Gaussian is 85. Figure~\ref{genev2} shows the
generated events, the PDE and the fitted curve for this case.

Let us note that it is possible to make a cumulative function from the
solid histogram in figure~\ref{fitlike} and estimate the probability
that ${\cal NLLR}$ exceeds the observed value, just as we do with
$\chi^2$ tests. We have also performed a binned $\chi^2$ fit to an
exponential over the same histograms, with the data in the fiducial
region binned over 41 bins. The resulting value of $\chi^2$ for 39
degrees of freedom are shown in the last column in
table~\ref{tab1}. At the 3$\sigma$ point for the unbinned method, the
binned method yields a $\chi^2$ of 42 over 39 degrees of freedom,
which may be considered a good fit. This implies that the unbinned
method may have more discriminating power against bad fits than the
binned one. It is worth noting however that the binned fit is over two
parameters (the number of events and the slope) whereas the unbinned fit
being considered here is only over a single parameter, namely the slope.

\begin{figure}[tbh!]
\centerline{\includegraphics[width=4.0in]{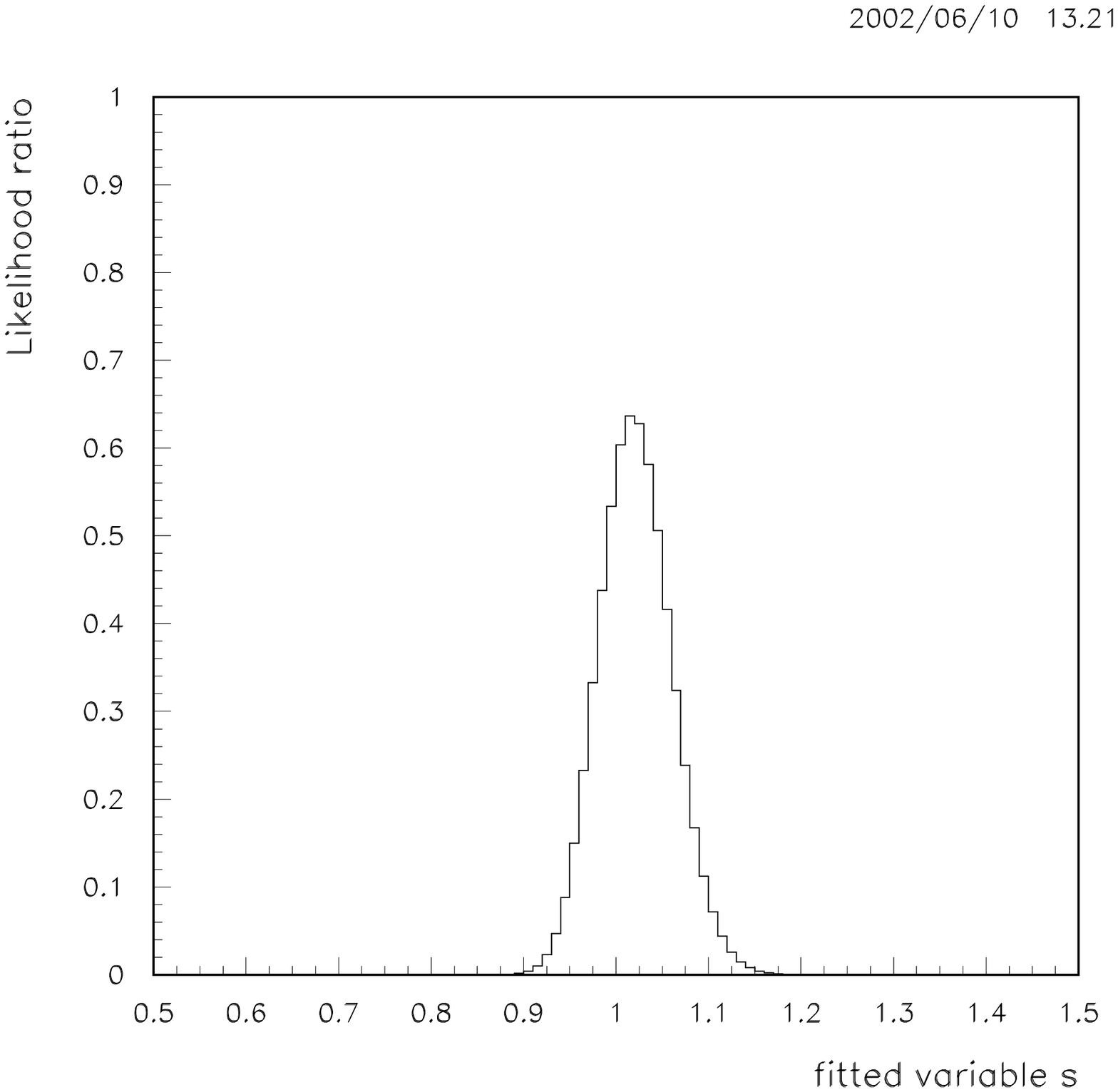}}
\caption[Fitted negative log likelihood ratio]
{Figure shows the  likelihood ratio ${\cal L_R}=\frac{P(c|s)}{P(c)}$ as a 
function of the fitted parameter $s$. 
The maximum likelihood point is at $s=1.019$.
\label{fitmax}}
\end{figure}

\begin{table}[bht!]
\caption[Fit results for various samples]{
\label{tab1}}
\centering\leavevmode
\begin{tabular}{|c|c|c|c|}
\hline
Number of & Unbinned fit & Unbinned fit& Binned fit $\chi^2$ \\
Gaussian events & ${\cal NLLR}$ & $N\sigma$ & 39 d.o.f.\\
\hline
500 & 189. & 103 & 304 \\
250 & 58.6 & 31 & 125 \\
100 & 11.6 & 4.9& 48 \\
85 & 8.2 & 3.0 & 42 \\
75 & 6.3 & 1.9 & 38 \\
50 & 2.55 & -0.14 &30 \\
0  & 0.44 & -1.33 & 24 \\
\hline
\end{tabular}
\end{table}
We now can fit the exponential data (with no superimposed Gaussian
bumps) and compute the value of the likelihood ratio 
${\cal L_R} = \frac{P(c|s)}{P(c)}$  as a function of the
parameter $s$ about the maximum likelihood point.  
Figure~\ref{fitmax} shows this function, which has the
maximum value at $s=1.019$. Note, however, that ${\cal L_R}$, a dimensionless
quantity, is not the likelihood distribution of $s$, which has to have the dimensions of $1/s$.

\subsection{Determination of the {\it a~priori} Likelihood P(s)}
In order to obtain the likelihood distribution of $s$,
$P(s|{\bf c_n})$, we need to understand better $ P(s)$,
the ${\it a~priori}$ distribution of $s$. We are in a position to do
this, since we have identified $P(c)$ to be the distribution of data,
and P(s) and P(c) are linked by the equation
\begin{equation}
P(c) = \int P(c|s) \times P(s)ds
\label{normpc1}
\end{equation}
We are in the process of using the {\it a posteriori} 
information contained in the data $pdf$  $P(c)$ to 
infer the {\it a priori} function $P(s)$. 
Before we use equation~\ref{normpc1} to calculate P(s), let us make
the following observations. Using equation~\ref{eqnlr}, we can write
\begin{equation}
P(s|{\bf c_n}) = P(s)\times {\cal L_R} = P(s) \times
\frac{P({\bf c_n}|s)}{P({\bf c_n})}
\label{post}
\end{equation}

As we increase $n$, the number of events sampled, in the limit
$n\rightarrow\infty$ we expect the ${\it a~posteriori}$ probability
$P({s|\bf c_n})$ to tend towards the delta function $\delta (s-s^*)$
where $s^*$ is the true value of $s$. This is because $P({s|\bf c_n})$
is the likelihood distribution of $s$ and we expect to determine the
true value of $s$ with infinite precision in this limit. However, the
ratio $\frac{P({\bf c_n}|s)}{P({\bf c_n})}$ will tend towards unity in
this limit (for a good fit), since for each data point $c_k$, the
theoretical $pdf~P(c_k|s)$ and the data $pdf~P(c_k)$ will be close to
each other. The only way out of this is to allow $P(s)$ to depend
discretely on $n$ and let the distribution $P(s)\rightarrow
\delta(s-s^*)$ as $n\rightarrow \infty$. We can see the need for this
further, using equation~\ref{normpc1}. In the limit
$n\rightarrow\infty$, the data $pdf$ P(c) will have the form
$P(c|s^*)$, where $s^*$ is the true value of $s$, if it fits the
theoretical model. Then the only solution for equation~\ref{normpc1}
is again $P(s) = \delta(s-s^*)$.

To repeat, the only way out of this dilemma, is for the
${\it a~ priori}$ probability distribution $P(s)$ to be dependent on $n$
and tend towards a delta function as $n\rightarrow\infty$. If we are
solving a Bayes theorem problem for $n$ data points, then the ${\it
a~priori}$ function $P(s)$ for that problem will be written as $P_n(s)$
indicating that it comes from a discrete familiy of probability
distributions that depend on $n$. Nothing further is known about
$P_n(s)$ ${\it a~priori}$ except that it is a $pdf$ in $s$ and that
the $pdf$ depends discretely on $n$.

\subsubsection{Rewriting the Bayes Theorem Equations}

The Bayes theorem equations have to be re-written to take into 
account this change.
Equation~\ref{eqnlr} now becomes

\begin{equation}
{\cal L_R}_{,n} = \frac{P(s|{\bf c_n})}{P_n(s)} = \frac{P({\bf
c_n}|s)}{P({\bf c_n})}
\end{equation}
Equation~\ref{eqnbay1} remains as is and equation~\ref{eqnbay2} becomes
\begin{equation}
{\cal L_R}_{,n} = \frac{P(s|{\bf c_n})}{P_n(s)} =
\frac{P(s|c_1)}{P_1(s)}\times\frac{P(s|c_2)}{P_1(s)}\dots
\times\frac{P(s|c_n)}{P_1(s)}
\end{equation}
 where we have also added the subscript $n$ to the likelihood ratio
${\cal L_R}$ to indicate its dependence on $n$.  The recursive chain
rule can now be rewritten as
\begin{eqnarray}
{\cal L_R}_{,k} = \frac{P(s|{\bf c_k})}{P_k(s)} =
\prod_{i=1}^{i=k}\frac{P(c_i|s)}{P(c_i)}\\
{\cal L_R}_{,l} = \frac{P(s|{\bf c_l})}{P_l(s)} =
\prod_{i=1}^{i=l}\frac{P(c_i|s)}{P(c_i)}\\
{\cal L_R}_{,k+l} = {\cal L_R}_{,k}\times{\cal L_R}_{,l} = 
 \frac{P(s|{\bf c_k})}{P_k(s)}\times \frac{P(s|{\bf c_l})}{P_l(s)} = 
\frac{P(s|{\bf c_{k+l}})}{P_{k+l}(s)} =
\prod_{i=1}^{i=k+l}\frac{P(c_i|s)}{P(c_i)}\label{multeq}
\end{eqnarray}
where we have two sub-samples of $k$ and $l$ events which are
being combined to form a total number of $k+l$ events.
\subsubsection{An Ansatz for $P_n(s)$} 
The expression for $P(c)$ in
equation~\ref{normpc1} can be thought of as the theoretical prediction
for $P(c)$ and the $PDE$ estimator is the experimental measurement of
$P(c)$. Then, one can write,
\begin{eqnarray}
\frac{P^{pred}({\bf c_n})}{P^{exp}({\bf c_n})} = 
\int \frac{P({\bf c_n}|s)}{P^{PDE}({\bf c_n})} \times P_n(s) ds\\
= \int {\cal L_R}_{,n} (s) \times P_n(s) ds = \int P(s|{\bf c_n}) ds =
1
\label{inteqn}
\end{eqnarray}
with the last expression following from Bayes theorem. There are two
ways the last equation $\int P(s|{\bf c_n}) ds = 1$ can be
satisfied. Either the likelihood ratio ${\cal L_R}_{,n} (s)$ = 1 or if
\begin{equation}
  P_n(s) = \frac{1}{\int {\cal L_R}_{,n}(s)ds}\equiv
  \frac{1}{2\lambda}
\label{pseq}
\end{equation}
It is very difficult for ${\cal L_R}_{,n} (s)$ to equal unity even at
the maximum likelihood value, since the experimental $PDE$ estimator
in the denominator is subject to statistical
fluctuations. Equation~\ref{pseq}, however, gives us an expression for
the {\it a priori} likelihood $P_n(s)$.  $P_n(s)$ is the value of the {\it a priori } probability distribution at the true value of $s$.
Since $\int P_n(s) ds = 1$, we can satisfy this 
with a functional form for  $P_n(s)$ being  a 
step function $\theta(s|\mu)$ such that
\begin{eqnarray}
\mbox{with} \: s_1 = \mu-\lambda \\
\mbox{and} \: s_2 = \mu + \lambda \\
P_n(s)\equiv \theta(s|\mu) = 0 \:\mbox{if}\: s< s_1\: \mbox{or}\: s>s_2 \\
P_n(s) \equiv \theta(s|\mu) = \frac{1}{2\lambda}~\mbox{if}\: s_1 \le s \le s_2 
\end{eqnarray}
and the value of $\mu$, the mean of
the distribution, is totally unknown.  {\it Let us note that it is possible
to write down an equation such as equation~\ref{pseq} only due to the
fact that we are able to compute a dimensionless quantity such as the
likelihood ratio ${\cal L_R}_{,n}(s)$ and  this becomes possible only due to
the concept of  the $PDE$ estimator for $P(c)$ introduced in this
paper.} The integral in equation~\ref{pseq} has thus the dimensions of
$s$ giving $P_n(s)$ the dimensions of $s^{-1}$ as required.

 As $n$ increases, the value of $\lambda$ decreases, since the
 distribution $P({\bf c_n}|s)$ sharpens. This has the effect of
 narrowing $P_n(s)$ in accordance with the discussion above. 
It is important to realize that there is only {\it one} true value of s, and the 
 Bayesian {\it a priori} probability $P_n(s)$ refers to the value of $P_n(s)$ at 
that true value, which, according to our ansatz, equals $\frac{1}{2\lambda}$.
The data does not result from an admixture of probable values of $s$, but from a single 
{\it true} value of $s$. So Bayes theorem becomes,
\begin{equation}
 P(s|{\bf c_n})\times P({\bf c_n}) = P({\bf c_n}|s)\times P_n(s) = P({\bf c_n}|s)\times \frac{1}{2\lambda}
\end{equation}
yielding
\begin{equation}
 P(s|{\bf c_n}) = \frac{ P({\bf c_n}|s)}{ P({\bf c_n})}\times \frac{1}{2\lambda} = 
\frac{{\cal L_R}(s)}{\int {\cal L_R}(s) ds} = \frac{ P({\bf c_n}|s)}{\int  P({\bf c_n}|s) ds}
\label{endres}
\end{equation}
 The last equation in~\ref{endres} results from the fact that the
 $PDE$ estimator for $P(c)$ cancels both in the numerator and
 denominator.  Having obtained $P(s|{\bf c_n})$, one can proceed to
 calculate the statistical quantities associated with $s$, namely the
 mean, mode, median, variance, errors and limits, in a rigorous
 fashion.  We note here that $P(s|{\bf c_n})$ is obtainable only with
 the use of of Bayes theorem, and our ansatz for the Bayesian {\it a
 priori} likelihood $P(s)$.  The expression for $P(s|{\bf c_n})$, the
 {\it a posteriori} likelihood for $s$ does not depend on the $PDE$
 estimator of data, but only on the theoretical function $P({\bf
 c_n}|s)$ evaluated at the data points.  The evaluation of $P_n(s)$
 and the goodness of fit criteria both require the usage of the $PDE$
 estimator for the data $pdf$.

The ansatz for the {\it a priori} distribution for $P_n(s)$ assumes a
flat distribution in $P_n(s)$.  This flatness may not be invariant
under change of variables and the consequences of this needs further
investigation.  It is important to stress again that $P_n(s)$ in the
Bayes theorem equations is the value of the {\it a priori}
distribution, at the true value of $s$. The value of the function
at the unknown true value of $s$ is known to 
some statistical precision
$(=\frac{1}{2\lambda})$.
We then use this to calculate the {\it a posteriori} distribution 
$P(s|{\bf c_n})$ which gives information about the true value of $s$ to some sttistical precision.
Since we use the value of the
function $P_n(s)$ at the true value of $s$ only, we may not be
sensitive to the shape of $P_n(s)$. Let us also note that we
do not use the {\it a priori} distribution explicitly for any calculations, 
since the information about
the error of $s$ is contained in the normalized ${\cal
L_R}(s)$. Combining data from different datasets may be done by
multiplying likelihood ratios as shown in equation~\ref{multeq}.

\begin{figure}[tbh!]
\centerline{\includegraphics[width=4.0in]{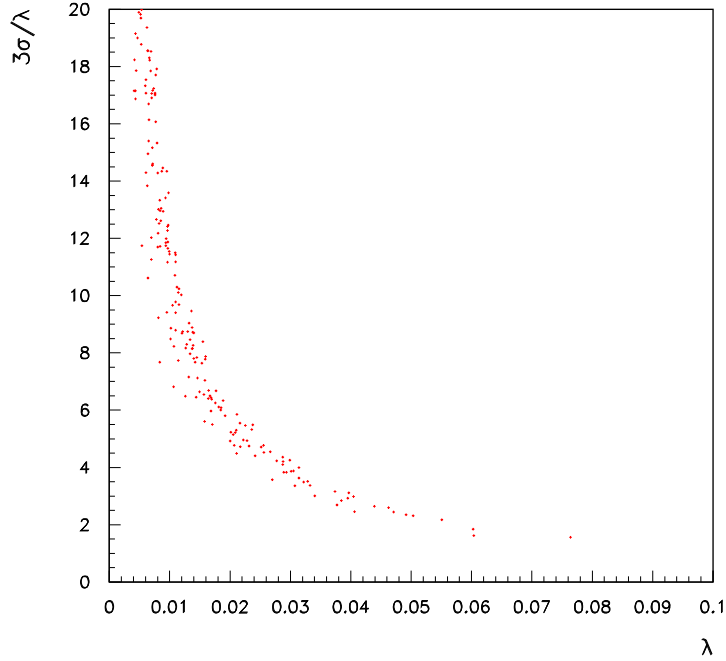}}
\caption[Integral of likelihood curve]
{Figure shows a scatter plot of $\lambda$, half the integral under the
likelihood curve vs.  $3\sigma/\lambda$, where $\sigma$ is the width of the 
likelihood distribution for 500 configurations.
\label{lambda}}
\end{figure}

 We note in passing that the the values of $\lambda$ are not large
 enough to span the width of the likelihood
 distribution. Figure~\ref{lambda} shows the correlation between
 $\lambda$ and the ratio (3$\sigma/\lambda$), where $\sigma$ is the
 $rms$  of the likelihood ratio distribution, for 500
 configurations ${\bf c_n}$ of 1000 events per configuration.  At no
 point does the ratio fall below unity, indicating that the likelihood
 curve is always broader than the step function $\theta(s|\mu)$.  We
 may not blindly use the step function as the {\it a priori}
 distribution, centered at the maximum likelihood value and do the
 integral in equation~\ref{inteqn}, since it will chop off the
 likelihood ratio curve in the tails. The step function can only be
 used after we feed it with a mean value $\mu$. The function in the
 integral in equation~\ref{inteqn} is in fact a constant which equals  the
 value of $P_n(s)$ at the true value of $s$. This value does {\it not}
 change as we change $s$ in the integral in equation~\ref{inteqn}.
 It is possible that the true value of $s$ is at the maximum likelihood point. 
It is also possible that it is at a value $3\sigma$ away, 
albeit with a reduced probability. The key point is that the true value of $s$ is
either at the maximum likelihood point {\it or} at any of the 
other values at which the likelihood is non-zero. They do not simultaneously have to be true.
Hence we can integrate over the whole likelihood distribution with 
$P_n(s) = \frac{1}{2\lambda}$ without worrying about falling off the 
edge of the step function. As one varies $s$, one is
testing mutually exclusive hypotheses that the value of $s$ under
consideration is the true value of $s$.

It is still instructive to see what happens when one supplies  a
distribution for the mean value of the step function. The following
section deals with the self-consistency of our expressions, when one
feeds in the {\it a posteriori} distribution $P(s|{\bf c_n})$ for the
mean value $\mu$ of the step function. 
\subsubsection{The Bootstrap}

If the mean value $\mu$ of the step function distribution has a
probability distribution $P(\mu)$, then one can write an expression
for the joint probability density of $\mu$ and $s$ as
\begin{equation}
  P(\mu)\times \theta(s|\mu) d\mu ds = \frac{P(\mu)}{2\lambda}d\mu ds
\end{equation}
\begin{figure}[tbh!]
\centerline{\includegraphics[width=4.0in]{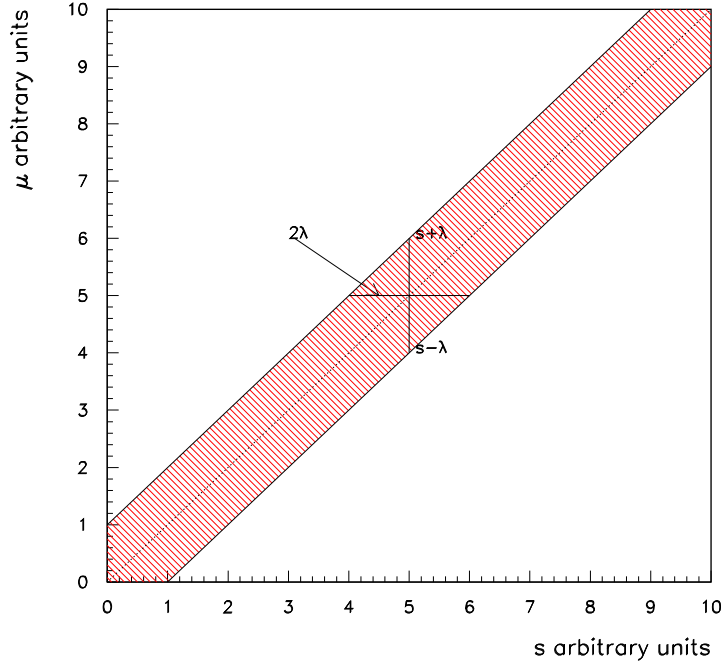}}
\caption[Bootstrap integral]
{The abscissa shows the variable $s$ and the ordinate the variable
$\mu$, the mean value of the $\theta$ function distribution. The
hatched region shows the area over which the probability distribution
for $s$ is non zero as a function of $\mu$.
\label{boot}}
\end{figure}
inside the shaded region in figure~\ref{boot} and is zero
outside. Integrating the above equation along the $s$ axis first
(fixed $\mu$), followed by integration along the $\mu$ axis yields
\begin{equation}
  \int_\mu P(\mu)\times d\mu \int_s \theta(s|\mu) ds = 1
\end{equation}
We can now reverse the order of integraton, doing the $\mu$
integration first, which yields
\begin{equation} \int_s \frac{1}{2\lambda} ds
\int_{s-\lambda}^{s+\lambda} P(\mu)d\mu = 1
\end{equation}
This can be re-written as
\begin{equation}
  \int_s \frac{1}{2\lambda}g(s)ds = 1
\label{eqint}
\end{equation}
where
\begin{equation}
  g(s) = \int_{s-\lambda}^{s+\lambda} P(\mu)d\mu
\end{equation}
These equations are true for any $P(\mu)$.  We need to supply a
$P(\mu)$, which is the probability distribution of the mean value of
the $\theta$ function. The obvious candidate for $P(\mu)$ is clearly
$P(s|{\bf c_n})$, the {\it a posteriori} distribution for the true
value of $s$. Since $\int g(s) ds = 2\lambda$, g(s) can be identified
with ${\cal L_R}(s)$, yielding
\begin{equation}
  g(s) \equiv {\cal L_R}(s) = \int_{s-\lambda}^{s+\lambda} P(s|{\bf
  c_n}) ds
\end{equation}
For small $\lambda$, we can Taylor expand the above integral yielding,
\begin{equation}
  {\cal L_R}(s) \approx 2\lambda \times P(s|{\bf c_n})
\end{equation}
yielding the desired result
\begin{equation}
   P(s|{\bf c_n}) \approx \frac{{\cal L_R}(s)}{2\lambda} = \frac{{\cal
   L_R}(s)}{\int{\cal L_R}(s)ds}
\label{boot1}
\end{equation}

Also, from equation~\ref{eqint}, we can identify $P(s|{\bf c_n})\equiv
\frac{g(s)}{2\lambda}$, since $\frac{g(s)}{2\lambda}$ is the
projection of the probability density for $s$ in figure~\ref{boot}
along the $s$ axis, computed after knowing ${\bf c_n}$.  This again
yields equation~\ref{boot1} and completes the bootstrap. 

To summarize the  arguments so far,
\begin{itemize}
\item We have made the ansatz that the {\it a priori} distribution for 
$P_n(s)$ is $\frac{1}{\int {\cal L_R}(s)ds}$. Such an ansatz gives us
a distribution $P_n(s)$ whose mean value is unknown but whose width
decreases with increased statistics. Both these properties qualify it
as a candidate for the {\it a priori} distribution. This step requires
a dimensionless ${\cal L_R}$ and is only possible by the use of the
experimental $PDE's$ for the goodness of fit test, introduced in this
paper.
\item We then supply it with a probability distribution for the mean
value, which is only known after we have analyzed ${\bf c_n}$.  The
candidate for the probability distribution for the mean $\mu$ is
$P(s|{\bf c_n})$, which is the {\it a posteriori} distribution for the
true value of $s$, and is the object of our quest. This is then used
to calculate the probabilty distribution of $s$.
\item This  yields the expression for  $P(s|{\bf c_n})$ of 
equation~\ref{boot1} as well as a probability density {\it a posteriori} 
for $s$ that is consistent with the same equation.
\end{itemize}
This results in
\begin{equation}
 P(s|{\bf c_n})  = \frac{{\cal L_R}_{,n} (s)}{\int {\cal L_R}_{,n}(s)ds} =
 \frac{P({\bf c_n}|s)}{\int P({\bf c_n}|s) ds}
\label{poster}
\end{equation}
for the {\it a posteriori} likelihood for $s$.

In multi-dimensional parameter space, with $\alpha$ being the dimension of the 
parameter vector $s$. the above  equations are generalized as follows
\begin{equation}
 P_n(s) = \frac{1}{\int {\cal L_R},n(s) ds}\equiv \frac{1}{(2\lambda)^\alpha}
\end{equation}
with integrals over $s$ being carried out over $\alpha$ dimensions.

\section{Towards an Analytic Theory of Unbinned Likelihood Goodness of Fit}
Because the likelihood ratio ${\cal L_R}(s)$ is invariant under transformation 
$c \rightarrow c^{'}$, one can use variables $c^{'}$ such that
\begin{equation}
 c^{'}(c) = \int_0^c P(c^{''}|s)dc^{''}
\end{equation}
This leads to probability distributions $P(c^{'}|s)$ such that
\begin{equation}
  P(c^{'}|s) = P(c|s) \times |\frac{dc}{dc^{'}}| = 1
\end{equation}
and with the limits of the variable $c^{'}$ being $ 0\: <\: c^{'}\:<\:1$.
These sets of transformations in multi-dimensions is known as 
the hypercube transformation. 
The number density is constant in the hypercube which implies 
that we are not sensitive to systematics associated with the 
smoothing parameter. 

The theoretical curve is a constant =1 in this scheme. The
experimental $PDE$ will also be close to being flat. The question to
answer is `` What is the distribution of the negative log likelihood
ratio ${\cal NLLR}$ that results from the statistical fluctuation of 
the $PDE$ in the hypercube''? We
leave this question to a subsequent paper.

\section{Conclusions}

We have introduced a technique for estimating goodness of fit in
unbinned likelihood fits by the use of probability density estimators
to obtain the {\it a priori} likelihood distribution of the data. In
addition to providing a measure of the goodness of fit in unbinned
likelihood fits for the first time, this approach enables us to obtain
expressions for the {\it a priori} likelihood distribution of the
theoretical parameters and hence to derive expressions for the {\it a
posteriori} likelihood distributions of the theoretical parameters. We
have shown that the {\it a priori} likelihood of the theoretical
parameters depends on the number $n$ of events being employed in the
problem. We have emphasized that the {\it a priori} likelihood is the value
of the probability distribution at the true value of $s$ and this does
not change as we change $s$, {\it a posteriori}, to calculate the
likelihood that $s$ is the true value.

The approach outlined in this paper permits the rigorous
calculation of errors in the fitted quantites. It makes unnecessary
the practice of ``guessing'' the {\it a priori} likelihood distributions of
parameters, a practice titled ``Bayesianism''. For the type of
problems considered here, the {\it a priori} likelihood distributions
can be computed.

The techniques detailed here are extensible to arbitrary dimensions, even 
though we have used a one-dimensional problem for illustrative purposes. In the process of using 
probability density estimators, we have developed an algorithm for iteratively 
improving the smoothing parameter as a function of local number density.
\section{Acknowledgements}
The author would like to thank British Rail for a particularly peaceful and 
appropriately long journey between Durham and London during which the main ideas in this 
paper were worked out.


\begin{thebibliography}{9}
\bibitem{fudge} 
After a modest amount of experimentation with
smoothing factors, we settled on the factor of 0.5 as being adequate to 
reproduce the exponential during this first phase of analysis. Later the 
smoothing function was introduced to improve the goodness of fit capability 
of the technique.
\bibitem{power} Empirically we found that a power of the order of -0.6 is
needed to provide sufficently large smoothing factors for large values of time.
We can in principle optimize this smoothing function further, 
but have not done so.
\end{thebibliography}
\end{document}